\documentstyle [12pt] {article}

\catcode`@=12
\begin{document}
\thispagestyle{empty}
\begin{flushright} UCRHEP-T97\\July 1992\
\end{flushright}
\vspace{0.3in}
\begin{center}
{\large \bf Gauge Model of Generation Nonuniversality Reexamined\\}
\vspace{0.8in}
{\bf Xiao-yuan Li\\}
\vspace{0.1in}
{\sl China Center of Advanced Science and Technology (CCAST)\\}
{\sl P. O. Box 8730, Beijing 100080, China\\}
{and\\}
{\sl Institute of Theoretical Physics\\}
{\sl Chinese Academy of Sciences\\}
{\sl P. O. Box 2735, Beijing 100080, China\\}
\vspace{0.2in}
{\bf Ernest Ma\\}
\vspace{0.1in}
{\sl Department of Physics\\}
{\sl University of California\\}
{\sl Riverside, California 92521\\}
\vspace{0.8in}
\end{center}
\begin{abstract}\
A gauge model of generation nonuniversality proposed in 1981 is reexamined
in the light of present precision electroweak measurements.  Two necessary
features of this model are that the $\tau$ lifetime should be longer than
predicted by the standard model and that the $\rho$ parameter measured at
the $Z$ peak should have an additional negative contribution.  Both trends
are noticeable in present data.  A future decisive test of this model would
be the discovery of new $W$ and $Z$ bosons with nearly degenerate masses of
a few $TeV$.
\end{abstract}

\newpage
\section{Introduction}

There are at least 3 generations of quarks and leptons, and each appears to
have identical gauge interactions.  In the context of the standard model,
each must then have the same representation content with respect to the
$SU(3) X SU(2) X U(1)$ gauge group.  Specifically, we have $(u_i,d_i)_L$,
$(\nu_i,l_i)_L$ as left-handed doublets, and $u_{iR},d_{iR},l_{iR}$ as
right-handed singlets, where $i=1,2,3$.  The quarks $u_i$ and $d_i$ are also
triplets under the color gauge group $SU(3)$, whereas the leptons $\nu_i$
and $l_i$ are singlets.  Different generations are distinguished by their
nontrivial Yukawa couplings to the one scalar doublet $(\phi^+,\phi^0)$ of
this model.  Masses and mixing come about as $\phi^0$ acquires a vacuum
expectation value $(vev)$, thereby breaking the $SU(2) X U(1)$ gauge symmetry
spontaneously.  The interaction universality of generations, which we may
call $e-\mu-\tau$ universality, is then a construction not from basic
principles but from the 2 inputs of good but not perfect experimental data
and the most straightforward way of implementing it theoretically.  As such,
we should not fail to question whether $e-\mu-\tau$ universality is truly
fundamental or not, and to seek an alternative understanding of this
phenomenon.

A lesson may be learned from our past and present understanding of strong
isospin.  This was thought to be a fundamental symmetry of the strong
interactions and was thought to be exact if electromagnetic and weak
interactions could be neglected.  For example, the proton-neutron mass
difference was thought to be entirely due to electromagnetism, but that
always gave the wrong sign.  We now know
of course that strong isospin, or flavor $SU(2)$ as it is often referred to
nowadays, is really only an {\it accidental} and {\it approximate} symmetry.
It comes about because the current quark masses $m_u$ and $m_d$ happen to
be much smaller than the interaction energy scale $\Lambda_{\rm QCD}$ of color
gauge $SU(3)$ which is the fundamental symmetry.  The proton-neutron mass
difference has an important contribution from $m_u-m_d < 0$ which is of order
$m_u$ itself in magnitude.  Furthermore, we can understand that since $m_s$
is not as small, flavor $SU(3)$ is even less exact.

Perhaps the well-tested $e-\mu$ universality and the less-well-tested
$e-\mu-\tau$ universality are also accidental and approximate symmetries
in analogy to flavor $SU(2)$ and $SU(3)$, and owe their existence to certain
mass-scale inequalities yet to be discovered.  We believe this to be a
theoretically attractive possibility and have proposed a specific
model\cite{lm81,mlt88,mn88,lm92} over 10 years ago to implement it.
In Section 2 we describe the model once again for the convenience of the
reader.  In Section 3 we take the limit of $e-\mu$ universality but not
$e-\mu-\tau$ universality, and show how the $\tau$ lifetime must necessarily
be longer than predicted by the standard model.  In Section 4 we show how
the observed $W$ and $Z$ bosons should differ from those of the standard
model, and show that in particular the $\rho$ parameter measured at the $Z$
peak should have an additional negative contribution.  In Section 5 we
constrain our model parameters against the present experimental data and
obtain bounds on the mass of the top quark from radiative corrections.  In
Section 6 we discuss the prediction that there should be new $W$ and $Z$
bosons with nearly degenerate masses of a few $TeV$ as a future decisive
test of this model.  Finally in Section 7, there are some concluding
remarks.

\section{The Model}

The electroweak gauge group is $U(1)~X~SU(2)_1~X~SU(2)_2~X~SU(2)_3$.  The
respective gauge couplings are $g_0,g_1,g_2,g_3$.  The left-handed quarks
\begin{equation}
\left[ \begin{array} {c} u' \\ d' \end{array} \right]_L ~,~ \left[
\begin{array} {c} c' \\ s' \end{array} \right]_L ~,~ \left[ \begin{array}
{c} t' \\ b' \end{array} \right]_L ~,
\end{equation}
and leptons
\begin{equation}
\left[ \begin{array} {c} \nu_e \\ e \end{array} \right]_L ~,~ \left[
\begin{array} {c} \nu_\mu \\ \mu \end{array} \right]_L ~,~ \left[
\begin{array} {c} \nu_\tau \\ \tau \end{array} \right]_L ~,
\end{equation}
are doublets coupling to separate $SU(2)$'s according to
\begin{equation}
Q = Y + I_3^{(1)} + I_3^{(2)} + I_3^{(3)}
\end{equation}
The right-handed quarks and leptons are singlets coupling only to $U(1)$.
The quark states $u',d',etc.$, are not mass eigenstates, but are related to
them $(u,d,etc.)$ by unitary transformations:
\begin{equation}
\left[ \begin{array} {c} u' \\ c' \\ t' \end{array} \right] = ~{\cal U} ~
\left[ \begin{array} {c} u \\ c \\ t \end{array} \right] ~,~~~ \left[
\begin{array} {c} d' \\ s' \\ b' \end{array} \right] = ~{\cal D} ~ \left[
\begin{array} {c} d \\ s \\ b \end{array} \right] ~.
\end{equation}
The lepton states are assumed to be mass eigenstates for simplicity.  The
scalar sector consists of 3 $SU(2)_i$ doublets
\begin{equation}
\Phi_i = \left[ \begin{array} {c} \phi_i^+ \\ \phi_i^0 \end{array} \right]
{}~,~~~ i = 1,2,3,
\end{equation}
and 3 self-dual $SU(2)_j~X~SU(2)_k$ bidoublets
\begin{equation}
\eta_i = {1 \over \sqrt 2} \left[ \begin{array} {c@{\quad}c} \eta_i^0 &
-\eta_i^+ \\ \eta_i^- & \overline {\eta}_i^0 \end{array} \right]
\end{equation}
such that $\eta_i = \tau_2 \eta_i^* \tau_2$.  Each column of $\eta_i$ is a
doublet under $SU(2)_j$ and each row is a doublet under $SU(2)_k$.  The
indices $i,j,k$ are cyclic permutations of 1,2,3.

The $vev$'s of $\phi_i^0$ are denoted by $v_i$ and those of $\eta_i^0$ by
$u_i$.\cite{mn88}
Since there are 3 $SU(2)$'s, there will be 3 sets of $W$ and $Z$ bosons.  The
$3X3$ mass-squared matrix of the $W$'s is given by
\begin{equation}
{\cal M}_W^2 = {1 \over 2} \left[ \begin{array} {c@{\quad}c@{\quad}c}
g_1^2 (v_1^2 + u_2^2 + u_3^2) & -g_1 g_2 u_3^2 & -g_1 g_3 u_2^2 \\
-g_1 g_2 u_3^2 & g_2^2 (v_2^2 + u_3^2 + u_1^2) & -g_2 g_3 u_1^2 \\
-g_1 g_3 u_2^2 & -g_2 g_3 u_1^2 & g_3^2 (v_3^2 + u_1^2 + u_2^2) \end{array}
\right] ,
\end{equation}
whereas the corresponding $4X4$ mass-squared matrix of the $Z$'s and the
photon $A$ is obtained from the above by adding a fourth row and column with
diagonal entry $g_0^2 (v_1^2 + v_2^2 + v_3^2) / 2$ and off-diagonal entries
$-g_0 g_i v_i^2 / 2$.  Consider first the charged-current interactions of the
leptons:
\begin{equation}
{\cal H}_{\rm int} = {1 \over \sqrt 2} \sum_i g_i W_{i\mu}^+ \overline {\nu}_i
\gamma^\mu \left( {{1-\gamma_5} \over 2} \right) l_i + {\rm H.c.}
\end{equation}
It is easily seen that the state
\begin{equation}
W = g_{123} (g_1^{-1} W_1 + g_2^{-1} W_2 + g_3^{-1} W_3),
\end{equation}
where
\begin{equation}
g_{123}^{-2} \equiv g_1^{-2} + g_2^{-2} + g_3^{-2},
\end{equation}
couples exactly to
\begin{equation}
\sum_i \overline {\nu}_i \gamma^\mu \left( {{1-\gamma_5} \over 2} \right) l_i~,
\end{equation}
with coupling $g_{123}$.  Hence if $W$ is indeed the lightest mass eigenstate
of Eq.(7) by far, then $e-\mu-\tau$ charged-current universality is realized
as a low-energy approximation.  A little algebra will verify that the
requirement for this to happen is simply that the $v_i$'s be much smaller
than the $u_i$'s, i.e. a mass-scale inequality as discussed before.  The
effective four-fermion interactions among the 3 lepton generations resulting
from the exchange of all 3 $W$'s of Eq.(8) are now characterized by a $3X3$
matrix
\begin{equation}
\left( {{4 G_F} \over \sqrt 2} \right)_{ij} = {1 \over 2} g_i g_j
({\cal M}_W^{-2})_{ij}
\end{equation}
which becomes $(v_1^2 + v_2^2 + v_3^2)^{-1}$ for all $i,j$ in the
universality limit.

Consider now the neutral-current interactions of the leptons:
\begin{eqnarray}
{\cal H_{\rm int}} & = & {1 \over 2} \sum_i (g_i W_{i\mu}^0 - g_0 B_\mu)
\left[ \overline {\nu}_i \gamma^\mu \left( {{1-\gamma_5} \over 2} \right)
\nu_i - \overline {l}_i \gamma^\mu \left( {{1-\gamma_5} \over 2} \right) l_i
\right] \nonumber \\ & - & g_0 B_\mu \sum_i \overline {l}_i \gamma^\mu l_i~,
\end{eqnarray}
where $B$ is the $U(1)$ gauge boson.  Now the photon $A$ is given by
\begin{equation}
A = e(g_0^{-1} B + g_1^{-1} W_1^0 + g_2^{-1} W_2^0 + g_3^{-1} W_3^0),
\end{equation}
where
\begin{equation}
e^{-2} \equiv g_0^{-2} + g_{123}^{-2},
\end{equation}
hence $g_i W_{i\mu}^0 - g_0 B_\mu$ is orthogonal to it as it must because
the neutrinos have no electric charge.  The effective four-fermion
interactions involving the neutrinos are then characterized by the coupling
matrix
\begin{equation}
\left( {{4 G_F'} \over \sqrt 2} \right)_{ij} = {1 \over 2} \left[ g_i g_j
({\cal M}_Z^{-2})_{ij} - g_i g_0 ({\cal M}_Z^{-2})_{i0} - g_0 g_j
({\cal M}_Z^{-2})_{0j} + g_0^2 ({\cal M}_Z^{-2})_{00} \right]
\end{equation}
which turns out to be {\it identical} to Eq.(12).  Note that the $4X4$ matrix
${\cal M}_Z^2$ does not really have an inverse because the photon is massless.
However we can simply give it an artificial mass $m_A$ and then do the
inversion.  The above expression is guaranteed to be nonsingular in $m_A$
and the desired result is then obtained by setting $m_A$ to zero.  The
equality of $(G_F)_{ij}$ and $(G_F')_{ij}$ in this model is the natural
generalization of the same relationship in the standard model.  The
underlying reason in both cases is the fact that only scalar {\it doublets}
have been used.  In the universality limit, the state
\begin{equation}
Z = e[g_{123}^{-1} B - g_0^{-1} g_{123} (g_1^{-1} W_1^0 + g_2^{-1} W_2^0 +
g_3^{-1} W_3^0)]
\end{equation}
becomes the lightest of the 3 $Z$'s by far and we recover all the predictions
of the standard model.

\section{Approach to $e-\mu-\tau$ Nonuniversality}

Since $e-\mu$ universality is more precisely verified than $e-\mu-\tau$
universality experimentally, we will assume for simplicity that the former
is exact in what follows.  This corresponds to taking the limit $u_3
\rightarrow \infty$ so that one set of $W$ and $Z$ bosons becomes very heavy
and decouples from the rest.  The resulting model is effectively
$U(1)~X~SU(2)_{12}~X~SU(2)_3$ with couplings $g_0, g_{12}, g_3$
respectively, where
\begin{equation}
g_{12}^{-2} \equiv g_1^{-2} + g_2^{-2},
\end{equation}
and the first 2 generations coupling to $SU(2)_{12}$ but the third coupling
to $SU(2)_3$.  The scalar sector is simplified to consist of only 2 doublets
and 1 bidoublet, with $vev$'s $v_{12} \equiv (v_1^2 + v_2^2)^{1/2}$, $v_3$,
and $u_{12} \equiv (u_1^2 + u_2^2)^{1/2}$ respectively.  In this limit, our
model closely resembles an earlier model\cite{bwm82} which assigned all 3
fermion generations to one of the $SU(2)$'s.  Our results will then differ
only if the third generation is involved.

Let us define
\begin{equation}
\xi \equiv 1 + {v_3^2 \over u_{12}^2},~~r \equiv {v_{12}^2 \over v_3^2},~~
y \equiv {g_{123}^2 \over g_3^2},
\end{equation}
then $G_F$, simplified to a $2X2$ matrix, is given by
\begin{equation}
{{4 G_F} \over \sqrt 2} = { \xi \over {v_3^2 (1 + r \xi)}} \left[
\begin{array} {c@{\quad}c} 1 & \xi^{-1} \\ \xi^{-1} & \xi^{-1}
(1\!+\!r(\xi\!-\!1)) \end{array} \right].
\end{equation}
This shows clearly that the rates for $\tau \rightarrow e \overline {\nu}_e
\nu_\tau$ and $\tau \rightarrow \mu \overline {\nu}_\mu \nu_\tau$ should be
smaller by the factor $\xi^{-2}$ than would be obtained from $e-\mu-\tau$
universality.  Experimentally, these rates are not directly measured, only
the branching fractions are; hence an independent measurement of the $\tau$
lifetime is also needed.  Up to a few months ago, the experimental
situation\cite{r91} was that a discrepancy existed at the level of
$2.3\sigma$, corresponding to $\xi-1 = 0.027 \pm 0.012$.  Since then, a
new value of $m_\tau = 1776.9 \pm 0.4 \pm 0.3~MeV$ has been announced by
the BES Collaboration\cite{z92} which is several $MeV$ lower than the
previously accepted value.  As the $\tau$ lifetime is proportional to
$m_\tau^{-5}$, this does reduce the size of the discrepancy but is not
enough to remove it.  New LEP data on the $\tau$ lifetime and branching
fractions have also become available.  Consequently, we now have\cite{rr92}
\begin{equation}
\xi - 1 = 0.015 \pm 0.008.
\end{equation}
Note that whereas the central value of $\xi-1$ is now reduced, the error
bars have also been reduced because of more precise data.  The statistical
significance is now $1.8\sigma$, down from $2.3\sigma$ of a few months ago.
Note also that in our previous work, the limit $r=0$ was assumed for
simplicity.  However, this does not affect the $\tau$-lifetime discrepancy
which depends only on $\xi$.

Consider for now only the first and second generations.  Their effective
neutral-current interactions at low energies are given by
\begin{equation}
{\cal H}_{\rm int} = {{4 G_F} \over \sqrt 2} \left[ (j_L^{(3)} - s^2 j^{em})^2
+ C (j^{em})^2 \right],
\end{equation}
where
\begin{equation}
s^2 = 1 - {e^2 \over g_0^2} - {e^2 \over g_3^2} \left( 1 - {1 \over \xi}
\right) = {e^2 \over g_{123}^2} \left[ 1 - y \left( 1 - {1 \over \xi}
\right) \right],
\end{equation}
and
\begin{equation}
C = {e^4 \over g_3^4} \left( 1 - {1 \over \xi} \right) \left( {1 \over \xi}
+ r \right) \simeq (\xi - 1) s^4 y^2 (1 + r).
\end{equation}
The standard-model limit is clearly obtained for $\xi=1$, in which case
$s^2=\sin^2 \theta_W$ and $C=0$.  Note that $s^2$ is also independent of $r$.
The $C$ term does not affect neutrino interactions at all and although it
does contribute in principle to, for example, $e^+e^- \rightarrow \mu^+\mu^-$
differential cross sections, it is highly suppressed numerically.

For completeness, we now write down the effective neutral-current interactions
at low energies which involve the third generation:
\begin{eqnarray}
{\cal H}_{\rm int} & = & {{4 G_F} \over \sqrt 2} \{~\xi^{-1} [1+r(\xi-1)]
j_L^{(3)}(\tau) j_L^{(3)}(\tau) + 2 \xi^{-1} j_L^{(3)}(\tau) j_L^{(3)}(e,\mu)
\nonumber \\ & - & 2 s^2 \left[ {{1+ry(\xi-1)} \over {\xi-y(\xi-1)}} \right]
j_L^{(3)}(\tau) j^{em}(e,\mu,\tau) - 2 s^2 j^{em}(\tau) j_L^{(3)}(e,\mu)
\nonumber \\ & + & s^4 \left[ 1 + {{(\xi-1) y^2 (1+r\xi)} \over
{[\xi-y(\xi-1)]^2}} \right] j^{em}(\tau)~(j^{em}(\tau)+2j^{em}(e,\mu))~\}.
\end{eqnarray}
The second term above tells us that the $e^+e^- \rightarrow \tau^+\tau^-$
and $e^+e^- \rightarrow \overline {b} b$ forward-backward asymmetries at
low energies should be reduced\cite{mlt88} by the factor $\xi^{-1}$
relative to the standard-model predictions.  However, present
data\cite{maki91} are not precise enough to determine the small deviation
indicated by Eq.(21).  To discover any possible additional evidence for
$e-\mu-\tau$ nonuniversality and in particular to determine the parameters
$y$ and $r$, we need then to consider the properties of the observed $W$
and $Z$ bosons.

\section{The $W$ and $Z$ Bosons}

Let us define
\begin{equation}
W_1^\pm \equiv g_{123} (g_{12}^{-1} W_{12}^\pm + g_3^{-1} W_3^\pm), ~W_2^\pm
\equiv g_{123} (g_3^{-1} W_{12}^\pm - g_{12}^{-1} W_3^\pm),
\end{equation}
then the $2X2$ mass-squared matrix in the basis $W_{1,2}^\pm$ is given by
\begin{equation}
{\cal M}_W^2 = {1 \over 2} g_{123}^2 v_3^2 \left[ \begin{array} {c@{\quad}c}
1+r & - (1\!-\!y)^{1 \over 2} y^{-{1 \over 2}} (1\!-\!ry(1\!-\!y)^{-1}) \\
- (1\!-\!y)^{1 \over 2} y^{-{1 \over 2}} (1\!-\!ry(1\!-\!y)^{-1}) &
y^{-1} (1\!-\!y)^{-1} (\xi\!-\!1)^{-1} + \Delta \end{array} \right],
\end{equation}
where $\Delta = y^{-1} (1-y) (1+ry^2(1-y)^{-2})$.  Similarly, for
\begin{equation}
Z_1 \equiv e [ g_{123}^{-1} B - g_0^{-1} g_{123} ( g_{12}^{-1} W_{12}^0 +
g_3^{-1} W_3^0)], ~Z_2 \equiv g_{123} (g_3^{-1} W_{12}^0 - g_{12}^{-1} W_3^0),
\end{equation}
we have
\begin{equation}
{\cal M}_Z^2 = \left[ \begin{array} {c@{\quad}c} e^{-2} g_0^2
({\cal M}_W^2)_{11} & - e^{-1} g_0 ({\cal M}_W^2)_{12} \\ - e^{-1} g_0
({\cal M}_W^2)_{21} & ({\cal M}_W^2)_{22} \end{array} \right].
\end{equation}
Note that Eqs.(27) and (29) reduce to Eqs.(33) and (44) of Ref.\cite{mn88}
in the limit $r=0$ as they should.  On the other hand, if $r=y^{-1}(1-y)$
instead, then both ${\cal M}_W^2$ and ${\cal M}_Z^2$ are diagonal and the
observed $W$ and $Z$ bosons are in fact $W_1$ and $Z_1$, each of which
couples universally to all 3 generations.  Hence it is entirely possible
for low-energy phenomena to exhibit $e-\mu-\tau$ nonuniversality, but not in
the observed $W$ and $Z$ bosons.  This is an important feature of our model
which was not previously recognized, and enables us to accommodate the
apparent universality in the precision measurements of the leptonic partial
widths of the $Z$ boson, to be discussed below.

In general, ${\cal M}_W^2$ and ${\cal M}_Z^2$ are not diagonal and there is
mixing between $W_1$ and $W_2$, and between $Z_1$ and $Z_2$.  Hence the
observed $W$ and $Z$ bosons are
\begin{equation}
W = W_1 \cos \theta + W_2 \sin \theta, ~Z = Z_1 \cos \varphi + Z_2 \sin
\varphi,
\end{equation}
where
\begin{equation}
\tan \theta \simeq (\xi - 1) y^{1 \over 2} (1 - y)^{3 \over 2}
(1 - ry (1 - y)^{-1}),
\end{equation}
and
\begin{equation}
\tan \varphi \simeq - e^{-1} g_0 \tan \theta \simeq - (1 - s^2)^{-{1 \over 2}}
\tan \theta.
\end{equation}
Their masses are then given by
\begin{equation}
M_W^2 \simeq {1 \over 2} g_{123}^2 v_3^2 (1+r) \left[ 1 - (\xi-1) {{(1-y)^2}
\over {1+r}} \left( 1 - {{ry} \over {1-y}} \right)^2 \right]
\end{equation}
and
\begin{equation}
M_Z^2 \simeq {g_0^2 \over e^2} M_W^2.
\end{equation}
{}From neutrino data, the parameter $s^2$ is extracted according to Eq.(22).
If we interpret this as the $\sin^2 \theta_W$ of the standard model,
the predicted masses-squared of the $W$ and $Z$ bosons would be
\begin{equation}
\mu_W^2 = {e^2 \over {4 \sqrt {2} G_F s^2}} \simeq {1 \over 2} g_{123}^2 v_3^2
(1 + r) \left[ 1 - (\xi - 1) {{1 - y} \over {1 + r}} \left( 1 - {{ry} \over
{1 - y}} \right) \right]
\end{equation}
and
\begin{equation}
\mu_Z^2 = {\mu_W^2 \over {1-s^2}} \simeq {g_0^2 \over e^2} \mu_W^2 \left[
1 - (\xi - 1) {{s^2 y} \over {1 - s^2}} \right].
\end{equation}
Hence
\begin{equation}
{M_W^2 \over \mu_W^2} \simeq 1 + (\xi - 1) y (1 - y) \left( 1 - {{ry} \over
{1 - y}} \right)
\end{equation}
and
\begin{equation}
{M_Z^2 \over \mu_Z^2} \simeq {M_W^2 \over \mu_W^2} + (\xi - 1) {{s^2 y}
\over {1 - s^2}}.
\end{equation}
If $r=0$, then the observed $W$ and $Z$ bosons should be heavier than
predicted by a small amount.  If $r=y^{-1}(1-y)$ so that $W$ and $Z$
exhibit $e-\mu-\tau$ universality, then only the $Z$ boson should be
heavier.

Consider the interactions of $Z_1$ and $Z_2$.
\begin{eqnarray}
{\cal H}_{\rm int} & = & - {{g_0 g_{123}} \over e} Z_1 \left[ j_L^{(3)} -
{e^2 \over g_{123}^2} j^{em} \right] \nonumber \\ & ~~ & + ~g_{123} Z_2
\left[ \left( {y \over {1-y}} \right)^{1 \over 2} j_L^{(3)}(e,\mu) - \left(
{{1-y} \over y} \right)^{1 \over 2} j_L^{(3)}(\tau) \right] \nonumber \\
& \simeq & - g_{eff} Z \left[ j_L^{(3)}(e,\mu) - {{e g_0} \over {g_{eff}
g_{123}}} j^{em}(e,\mu) \right] + ~...
\end{eqnarray}
where
\begin{equation}
g_{eff} \simeq {{g_0 g_{123}} \over e} \left[ 1 + (\xi-1) y (1-y) \left(
1 - {{ry} \over {1-y}} \right) \right].
\end{equation}
Now the widths and forward-backward asymmetries of $Z \rightarrow l \overline
{l} ~(l = e, \mu, \tau)$ are given by
\begin{equation}
\Gamma_l = {{G_F M_Z^3} \over {24 \sqrt {2} \pi}} \left( 1 + {{3 \alpha} \over
{4 \pi}} \right) \rho_l \left[ 1 + (1 - 4 \sin^2 \theta_l)^2 \right],
\end{equation}
and
\begin{equation}
A_{FB}^{l}(M_Z^2) \simeq 3 (1 - 4 \sin^2 \theta_l)^2.
\end{equation}
Therefore, we have
\begin{equation}
\rho_{e,\mu} = {g_{eff}^2 \over {4 \sqrt {2} G_F M_Z^2}} \simeq 1 - (\xi-1)
y^2 (1+r),
\end{equation}
and
\begin{equation}
\sin^2 \theta_{e,\mu} \simeq s^2 [ 1 + (\xi-1) y^2 (1+r)].
\end{equation}
We see thus a necessarily negative contribution to $\rho_{e,\mu}$ in this
model, which can compensate against the well-known positive radiative
contribution of the $t$ quark, to be discussed in the next section.
Similarly, we can extract from ${\cal H}_{\rm int}$ of Eq.(39) the
corresponding parameters for the third generation:
\begin{equation}
\rho_\tau \simeq 1 - (\xi-1) [2 - y(2-y)(1+r)],
\end{equation}
and
\begin{equation}
\sin^2 \theta_\tau \simeq s^2 [1+(\xi-1)(1-y(1-y)(1+r))],
\end{equation}
so that
\begin{equation}
{\Gamma_\tau \over \Gamma_{e,\mu}} \simeq 1 - 2(\xi-1) {{(1-y)(1-2s^2)} \over
{1-4s^2+8s^4}} \left( 1 - {{ry} \over {1-y}} \right).
\end{equation}
In the limit $r=0$, it is clear that we must have $\Gamma_\tau <
\Gamma_{e,\mu}$,\cite{lm92} but unlike the conditions $\xi > 1$ and
$\rho_{e,\mu} < 1$, this should not be thought of as a necessary
prediction of our model.  For neutrinos, the ratio $\Gamma_{\nu_\tau} /
\Gamma_{\nu_e,\nu_\mu}$ is obtained from Eq.(47) by setting $s^2$ to zero.

Since our model has small tree-level corrections to the experimentally
measured quantities $M_Z$, $\rho_l$, $\sin^2 \theta_l$, and $M_W$, we can
also express its deviation from the standard model in terms of additional
contributions to the "oblique" parameters $\epsilon_{1,2,3}$\cite{abj92}
even in the absence of radiative corrections.  For $e$ and $\mu$, we have
\begin{eqnarray}
\Delta \epsilon_1 = \Delta \epsilon_2 = - (\xi-1) y^2 (1+r), \nonumber \\
\Delta \epsilon_3 = (\xi-1) y [1-y(1+r)],
\end{eqnarray}
and for $\tau$,
\begin{eqnarray}
\Delta \epsilon_1 = \Delta \epsilon_2 = - (\xi-1) [2-y(2-y)(1+r)],
\nonumber \\ \Delta \epsilon_3 = - (\xi-1) (1-y) [1-y(1+r)].
\end{eqnarray}

In the $W$ sector, the analog of Eq.(39) is
\begin{eqnarray}
{\cal H}_{\rm int} & = & \sqrt 2 g_{123} W_1^+ j_L^+(e,\mu,\tau) \nonumber \\
& + & \sqrt 2 g_{123} W_2^+ \left[ \left( {y \over {1-y}} \right)^{1 \over 2}
j_L^+(e,\mu) - \left( {{1-y} \over y} \right)^{1 \over 2} j_L^+(\tau)
\right] + {\rm H.c.}
\end{eqnarray}
where
\begin{equation}
j_L^+(l) = {1 \over 2} \overline {\nu}_l \gamma \left( {{1-\gamma_5} \over 2}
\right) l.
\end{equation}
Using Eqs.(30) and (31), we then have
\begin{equation}
{{\Gamma (W^+ \rightarrow \tau^+ \nu_\tau)} \over {\Gamma (W^+ \rightarrow
e^+ \nu_e, \mu^+ \nu_\mu)}} \simeq 1 - 2(\xi - 1) (1-y) \left(1 - {{ry} \over
{1-y}} \right),
\end{equation}
which is identical to the corresponding ratio for $Z \rightarrow \nu
\overline {\nu}$ as expected.

\section{Constraints from Present Data}

The mass of the $Z$ boson is now precisely measured experimentally:\cite{lep92}
\begin{equation}
M_Z = 91.175 \pm 0.021~GeV.
\end{equation}
{}From this and two other precisely measured quantities, it is useful to define
\begin{equation}
s_0^2 (1-s_0^2) \equiv { {\pi \alpha (M_Z^2)} \over {\sqrt {2} G_F M_Z^2} },
\end{equation}
where $G_F=1.16637X10^{-5}~GeV^{-2}$ is the Fermi coupling constant and
$\alpha (M_Z^2)$ is the fine-structure constant extrapolated from its value
at zero momentum transfer $(q^2=0)$ to $q^2=M_Z^2$, the most recent
precise determination being\cite{dfs91}
\begin{equation}
\alpha^{-1}(M_Z^2) = 127.9 \pm 0.2.
\end{equation}
Hence we obtain
\begin{equation}
s_0^2 = 0.2338 \pm 0.0005.
\end{equation}

To facilitate our model comparisons with data, let us define a new quantity
\begin{equation}
x \equiv y (1+r)
\end{equation}
and use it together with $y$ and $\xi-1$ in the following analysis.  Since
$r \geq 0$ by definition, we must always have $x \geq y$.  The special cases
of $r=0$ and $r=y^{-1} (1-y)$ correspond to $x=y$ and $x=1$ respectively.
We will be using as experimental inputs the measured values of $M_Z$,
$\rho_l$, and $\sin^2 \theta_l$ at LEP, as well as that of $\sin^2 \theta_W$
from neutrino data.  The dominant radiative correction will be assumed to be
from the $t$ quark.

To constrain $x$ and $y$ for a given value of $\xi-1$, we consider the
following quantities:
\begin{equation}
{\Gamma_\tau \over \Gamma_{e,\mu}} \simeq 1 - {{2(\xi-1)(1-2s_0^2)} \over
{1-4s_0^2+8s_0^4}} (1-x),
\end{equation}
\begin{eqnarray}
\rho_{eff} & \equiv & {1 \over 3} (\rho_e + \rho_\mu + \rho_\tau) \nonumber \\
& \simeq & 1 + \rho_{rad} - (\xi-1) \left[ {2 \over 3} (1-x) + xy \right],
\end{eqnarray}
\begin{eqnarray}
\sin^2 \theta_{eff} & \equiv & {1 \over 3} (\sin^2 \theta_e + \sin^2
\theta_\mu + \sin^2 \theta_\tau) \nonumber \\ & \simeq & s_0^2 \left\{ 1 -
\left( {{1-s_0^2} \over {1-2s_0^2}} \right) \rho_{rad} + (\xi-1) \left[
{1 \over 3} (1-x) + {{y(1-xs_0^2)} \over {1-2s_0^2}} \right] \right\} \nonumber
\\ & + & {\rm other~assumed~negligible~radiative~corrections},
\end{eqnarray}
\begin{equation}
{M_Z^2 \over \mu_Z^2} \simeq 1 + (\xi-1) y \left[ {1 \over {1-s_0^2}} - x
\right],
\end{equation}
and
\begin{equation}
\mu_Z = {{37.281~GeV} \over {\sin \theta_W \cos \theta_W \sqrt {1-\Delta r}}},
\end{equation}
where $\Delta r$ is the standard-model radiative correction to $\mu_Z$ in
the on-shell renormalization scheme.  The radiative correction $\rho_{rad}$
is assumed to be dominated by the $t$ quark:
\begin{equation}
\rho_{rad} \simeq {{3 \sqrt {2} G_F m_t^2} \over {16 \pi^2}},
\end{equation}
which contributes significantly to $\rho_{eff}$ and $\sin^2 \theta_{eff}$.

In addition to the published 1989 and 1990 LEP data,\cite{lep92} the
preliminary 1991 data are now available.\cite{nash92}
We combine them to find
\begin{eqnarray}
\Gamma_e &=& 83.37 \pm 0.32~MeV, \\
\Gamma_\mu &=& 83.59 \pm 0.49~MeV, \\
\Gamma_\tau &=& 83.38 \pm 0.60~MeV,
\end{eqnarray}
and
\begin{eqnarray}
\rho_{eff} &=& 0.9990 \pm 0.0032, \\
\sin^2 \theta_{eff} &=& 0.2322 \pm 0.0015.
\end{eqnarray}
We will also use
\begin{equation}
\sin^2 \theta_W = 0.231 \pm 0.006
\end{equation}
from neutrino data\cite{haidt90} and the values of $\Delta r$ as a function
of $m_t$ as given in Ref.\cite{jeger90}

First we note that given the values of $s_0^2$ and $\xi-1$, we can determine
$x$ from Eq.(58).  Actually, a small kinematic correction must be applied
here because $m_\tau$ is not as negligible as $m_e$ and $m_\mu$ compared
to $M_Z$.  This is equivalent to increasing the measured value of
$\Gamma_\tau$ in Eq.(66) by $0.19~MeV$ before dividing by $\Gamma_{e,\mu}$.
Therefore we find
\begin{equation}
{\Gamma_\tau \over \Gamma_{e,\mu}} = {{83.57 \pm 0.60} \over {83.44 \pm 0.27}}
= 1.0016 \pm 0.0079,
\end{equation}
which yields
\begin{equation}
1 - {0.0030 \over {\xi-1}} < x < 1 + {0.0045 \over {\xi-1}}
\end{equation}
as shown in Fig. 1.  For illustration, let us suppose $x=1$, then we have
from Eqs.(59) and (67),
\begin{equation}
\rho_{rad} - (\xi-1) y = -0.0010 \pm 0.0032~;
\end{equation}
from Eqs.(56), (60), and (68),
\begin{equation}
\rho_{rad} - (\xi-1) y = 0.0048 \pm 0.0047~;
\end{equation}
and from Eqs.(53), (61), (62), and (69),
\begin{equation}
\Delta r > 0.0417 - 0.292 (\xi-1) y.
\end{equation}
In the standard-model limit of $\xi-1=0$, these would correspond to 1$\sigma$
upper bounds of 84 $GeV$ and 174 $GeV$ on $m_t$ from Eqs.(72) and (73)
respectively, and an upper bound of 147 $GeV$ on $m_t$ if $m_H=100~GeV$ from
Eq.(74), whereas the present experimental lower bound\cite{cdf92} is 91 $GeV$.
In our model, $\xi-1$ ranges from 0.007 to 0.023 and $y$ ranges from 0 to 1,
hence the upper bounds on $m_t$ from Eqs.(72) and (73) are much higher,
{\it i.e.} 284 $GeV$ and 322 $GeV$ respectively.  On the other hand,
$\Delta r$ is not as sensitive to $m_t$ and Eq.(74) still requires
$m_t < 164~GeV$ if $m_H=100~GeV$.  This means that our model also prefers
a value of $m_t$ below 200 $GeV$.  The difference with the standard model
is that the most stringent constraint now comes from neutrino data and $M_Z$,
rather than from $\rho_l$.

Using Eqs.(21) and (71) to constrain $\xi-1$ and $x$, we now compare Eq.(59)
against Eq.(67), Eq.(60) against Eqs.(56) and (68), and $\Delta r$ from
Eq.(62) against Eqs.(53), (61), and (69), to constrain $y$ and $m_t$.
Keeping in mind that $x \geq y$ by definition, we search for the maximum
value of $m_t$ allowed within one standard deviation of all the data.
Our result is
\begin{equation}
m_t < 158~GeV.
\end{equation}
This upper bound is obtained at the low end of $\xi-1$, but it changes only
slightly to 155 $GeV$ at the high end.  To visualize how present data
constrain the $x$ and $y$ parameters of our model, let us consider 4 special
cases:  $m_t=150$ and 120 $GeV$, with $\xi-1=0.01$ and 0.02.  We show in
Figs. 2 to 5 the allowed regions in $x$ and $y$ for each case.  As a
function of $x$, the $\sin^2 \theta_l$ data always provide an upper bound on
$y$, but no useful lower bound, whereas the $\rho_l$ data always provide a
lower bound on $y$ and sometimes also an upper bound which is however less
restrictive than that of the $\sin^2 \theta_l$ data.  The neutrino data
provide a lower bound on $y$, which is also a steeply rising function of $x$.
At $m_t=120~GeV$, neutrino data do not restrict $x$ and $y$ beyond their
theoretically allowed ranges of $x \geq y$ and $0 \leq y \leq 1$.  As $m_t$
increases, the derived lower bound on $y$ becomes nontrivial and will
eventually eliminate the entire $x$ and $y$ parameter space.
For given values of $\xi-1$ and $x$, there is also a lower bound on $m_t$
as a function of $y$ from $\sin^2 \theta_l$ data, but it is not very useful
because it is actually less than the experimental lower bound of 91 $GeV$
for some allowed range of $y$.

In our discussion so far, we have assumed that $\rho_{rad}$ is dominated
by the $t$ quark, as given by Eq.(63).  This means that we have neglected
all the scalar contributions.  In the standard model, there is only one
physical Higgs boson and it does contribute to $\Delta r$ of Eq.(62).  In
converting a numerical bound on $\Delta r$ from Eq.(74) to a bound on $m_t$,
we have assumed $m_H=100~GeV$.  If we allow a larger $m_H$, then our upper
bound on $m_t$ goes up, but since the dependence on $m_H$ is only
logarithmic, a change from $m_H=100~GeV$ to 1000 $GeV$ will shift $m_t$
upward by less than 20 $GeV$.  In our model, there are many physical scalar
particles and they will certainly contribute to $\rho_{rad}$ and to
$\sin^2 \theta_{eff}$ as indicated in Eq.(60).  However, the structure of
our scalar sector is such that an automatic custodial $SU(2)$ symmetry is
present\cite{mn88,bwm82} so that $\rho_{rad}$ has no quadratic scalar mass
terms to one-loop order.  The residual contributions are all logarithmic in
these unknown scalar masses and may not be negligible, but they are not
expected to change our results in a drastic way.  The reason for this
custodial $SU(2)$ symmetry is that there is only one copy of each type of
scalar doublet.  If there were two copies of the same type of doublet,
Eq.(63) would not have been a good approximation, as is well-known in a
general two-scalar-doublet extension of the standard model.\cite{fmk92}

We have not used any of the $Z \rightarrow q \overline {q}$ data because
they have large QCD corrections and unless these are known very well, it
would be impossible to extract the small additional contributions of our
model.  At present, there is an apparent disagreement regarding the value
of $\alpha_S$ determined from event shapes which give $\alpha_S \simeq 0.12$,
and from partial widths which give $\alpha_S \simeq 0.14$.  As for other
available data such as $Z \rightarrow \nu \overline \nu$ as well as the
mass and partial widths of the $W$ boson, they are certainly consistent
with our model but are not precise enough at present to be of any practical
importance.

\section{Future Implications}

The effects due to nonuniversality are naturally very small at present
energies.  They will probably never be a decisive test of our model.
However, we also predict the existence of a second set of $W$ and $Z$
bosons with nearly degenerate masses approximately given by
\begin{equation}
M_{W'}^2 = M_{Z'}^2 = { {M_Z^2 (1-s_0^2)} \over {(\xi-1)(1-y)x} },
\end{equation}
as shown in Eqs.(27) and (29).  The reason for this mass degeneracy has
to do with the symmetry breaking pattern of $SU(2) X SU(2) \rightarrow SU(2)$.
The 3 gauge bosons which acquire mass at this level, {\it i.e.} $W_2^\pm$
and $Z_2$, transform as a global triplet under the unbroken $SU(2)$ and
must therefore have the same mass.  Since $W'$ and $Z'$ are identical to
$W_2$ and $Z_2$ except for small admixtures of $W_1$ and $Z_1$, their masses
have to be nearly degenerate.  To minimize this common mass, we need to
maximize $(\xi-1)(1-y)x$, subject to the constraints of present data.  We
find $\xi-1=0.023$, $y=0$, $x=0.974$, and $m_t=91~GeV$, corresponding to
a lower bound of 533 $GeV$ for $M_{W'}=M_{Z'}$.  There is no upper bound at
present, because the $y=1$ limit cannot be ruled out by the data.  However,
if we take the symmetry limit\cite{kuoetal} $g_1=g_2=g_3$ and $v_1=v_2=v_3$
so that $y=1/3$ and $x=1$, then we have an upper bound of about 1.17 $TeV$.
Hence it is not unreasonable to expect in general an upper bound of
no more than a few $TeV$.  These new vector gauge bosons should then be
within reach of discovery at future accelerators such as the SSC or LHC.

The interactions of our new $W'$ and $Z'$ bosons are already given in
Eqs.(50) and (39).  Their couplings are well approximated by those of
$W_2$ and $Z_2$, namely $g_{123} \sqrt {y} / \sqrt {1\!-\!y}$ to the first two
generations and $-g_{123} \sqrt {1\!-\!y} / \sqrt {y}$ to the third.  Since
$g_{123} \simeq e/\sin \theta_W$, the production of $W'$ and $Z'$ via
quark fusion should be substantial at hadronic colliders if there is
enough energy.  For example, with an integrated luminosity of 10 $fb^{-1}$
at the SSC, using SDC cuts, we expect\cite{rizzo} the order of 100 events
each for $e^+e^-$, $\mu^+\mu^-$, and $\tau^+\tau^-$ coming from $Z'$ if
$y=1/3$ and $M_{Z'}=3.5~TeV$.  Note also that the fermionic current which
couples to $Z'$ is predominantly left-handed, and that which couples to
$W'$ is completely left-handed.  This feature should help to distinguish
our model from other extensions of the standard model.\cite{hr92}

\section{Conclusion}

The universality of $e-\mu-\tau$ interactions at low energies may only be
an accidental, approximate symmetry analogous to that of flavor $SU(2)$
and $SU(3)$.  This was specifically realized by an electroweak gauge model
based on $U(1)~X~SU(2)_1~X~SU(2)_2~X~SU(2)_3$ we proposed in 1981.
To distinguish our model from the standard model, we must first have a
longer $\tau$ lifetime so that $\xi > 1$.  At present, this effect is at
the level of 1 or 2 percent, {\it i.e.} $\xi-1 = 0.015 \pm 0.008$.
In the future, the allowed range of $\xi-1$ may be narrowed as more precise
data on the $\tau$ lifetime and leptonic branching fractions become
available.  However, as long as the data are consistent with $\xi-1>0$
within their error bars, our model cannot be ruled out.  We should then
consider other possible effects such as $e-\mu-\tau$ nonuniversality in
$Z$ decay, which determines the combination $(\xi-1)(1-x)$ as shown in Eq.(58)
and Fig. 1.  Note that $Z$ decay may exhibit universality because $x=1$ but
$\xi-1$ may still be nonzero.  The values of $\rho_l$ and $\sin^2 \theta_l$
as well as $\sin^2 \theta_W$ from neutrino data depend also on $y$ and
$\rho_{rad}$, so they can be used to constrain $y$ and $m_t$ if we assume
Eq.(63).  Note that by definition, we must have $0 \leq y \leq 1$ and
$x \geq y$.

In this paper, we have shown that our model is completely consistent with all
present data and does better than the standard model in two specific
measurements, namely the $\tau$ lifetime and $\rho_l$.  Mainly because of
the measured value of $\sin^2 \theta_W$ from neutrino data, our model still
requires an upper bound on $m_t$ similar to that of the standard model.
Assuming that scalar contributions are negligible, we find $m_t < 158~GeV$
in order that all the data we consider are satisfied within one standard
deviation.  Once the $t$ quark is discovered and $m_t$ is known with some
accuracy, our model parameters $x$ and $y$ may be constrained as shown in
Figs. 2 to 5.  This will then lead to a decisive prediction of our model,
namely the existence of the $W'$ and $Z'$ bosons, with masses of the order
$TeV$ as given by Eq.(76). They are predicted to have the same mass, each
breaking $e-\mu-\tau$ universality as discussed in Section 6.

Whether or not $e-\mu-\tau$ universality is a fundamental symmetry is an
important question.  It should of course be decided experimentally, but
there are always practical limitations.  A case in point is CP as a
possible fundamental discrete symmetry.  We now know of course that it
is not exact, but the nonconserving effect is very small and we may be
considered lucky that the masses of the $K$ and $\pi$ mesons are such that
it has revealed itself to us experimentally.  Even if $e-\mu-\tau$ universality
is not exact, it may be very difficult to observe the small deviation.
However, in our specific realization of $e-\mu-\tau$ universality as an
accidental, approximate symmetry, there will be an unambiguous test in the
future at the SSC or LHC.
\vspace{0.3in}
\begin{center} {ACKNOWLEDGEMENT}
\end{center}

We thank Tom Rizzo for many important discussions.
This work was supported in part by the U. S. Department of Energy under
Contract No. DE-AT03-87ER40327.

\newpage
\bibliographystyle{unsrt}

\newpage
\begin{center}  {FIGURE CAPTIONS}
\end{center}

Fig. 1.  Plot of $x$ versus $\xi-1$ as allowed by Eqs.(21) and (71).

Fig. 2.  Constraints on $x$ and $y$ due to various data for
$m_t=150~GeV$ and $\xi-1=0.01$.

Fig. 3.  Constraints on $x$ and $y$ due to various data for
$m_t=150~GeV$ and $\xi-1=0.02$.

Fig. 4.  Constraints on $x$ and $y$ due to various data for
$m_t=120~GeV$ and $\xi-1=0.01$.

Fig. 5.  Constraints on $x$ and $y$ due to various data for
$m_t=120~GeV$ and $\xi-1=0.02$.

\end{document}